\documentclass[twocolumn,showpacs,preprintnumbers,amsmath,amssymb]{revtex4}

\usepackage{graphicx}
\usepackage{dcolumn}
\usepackage{bm}
\usepackage{amsmath}

\begin{document}
\title{Behaviors of susceptible-infected epidemics on scale-free networks with identical infectivity}
\author{Tao Zhou$^{1,2}$}
\email{zhutou@ustc.edu}
\author{Jian-Guo Liu$^{3}$}
\author{Wen-Jie Bai$^{4}$}
\author{Guanrong Chen$^{2}$}
\author{Bing-Hong Wang$^{1}$}
\email{bhwang@ustc.edu.cn}
\affiliation{%
$^{1}$Department of Modern Physics and Nonlinear Science Center,
University of Science and Technology of China, Anhui Hefei 230026,
PR China\\
$^{2}$ Department of Electronic Engineering, City University of Hong Kong, Hong Kong SAR, PR China\\
$^{3}$Institute of System Engineering, Dalian
University of
Technology, Dalian 116023, PR China\\
$^{4}$Department of Chemistry, University of Science and
Technology of China, Anhui Hefei 230026, PR China
}%

\date{\today}

\begin{abstract}
In this article, we proposed a susceptible-infected model with
identical infectivity, in which, at every time step, each node can
only contact a constant number of neighbors. We implemented this
model on scale-free networks, and found that the infected
population grows in an exponential form with the time scale
proportional to the spreading rate. Further more, by numerical
simulation, we demonstrated that the targeted immunization of the
present model is much less efficient than that of the standard
susceptible-infected model. Finally, we investigated a fast
spreading strategy when only local information is available.
Different from the extensively studied path finding strategy, the
strategy preferring small-degree nodes is more efficient than that
preferring large-degree nodes. Our results indicate the existence
of an essential relationship between network traffic and network
epidemic on scale-free networks.
\end{abstract}

\pacs{89.75.-k,89.75.Hc,87.23.Ge,05.70.Ln}

\maketitle

\section{Introduction}
Since the seminal works on the small-world phenomenon by Watts and
Strogatz \cite{Watts1998} and the scale-free property by
Barab\'asi and Albert \cite{Barabasi1999}, the studies of complex
networks have attracted a lot of interests within the physics
community \cite{Albert2002,Dorogovtsev2002}. One of the ultimate
goals of the current studies on complex networks is to understand
and explain the workings of the systems built upon them
\cite{Newman2003,Boccaletti2006}. The previous works about
epidemic spreading in scale-free networks present us with
completely new epidemic propagation scenarios that a highly
heterogeneous structure will lead to the absence of any epidemic
threshold (see the review papers \cite{Pastor2003,Zhou2006} and
the references therein). These works mainly concentrate on the
susceptible-infected-susceptible (SIS)
\cite{Pastor2001a,Pastor2001b} and susceptible-infected-removed
(SIR) \cite{May2001,Moreno2002} models. However, many real
epidemic processes can not be properly described by the above two
models. For example, in many technological communication networks,
each node not only acts as a communication source and sink, but
also forwards information to others
\cite{Tanenbaum1996,Krause2006}. In the process of broadcasting
\cite{Park2005,Harutyunyan2006}, each node can be in two discrete
states, either \emph{received} or \emph{unreceived}. A node in the
received state has received information and can forward it to
others like the infected individual in the epidemic process, while
a node in the unreceived state is similar to the susceptible one.
Since the node in the received state generally will not lose
information, the so-called susceptible-infected (SI) model is more
suitable for describing the above dynamical process. Another
typical situation where the SI model is more appropriate than SIS
and SIR models is the investigation of the dynamical behaviors in
the very early stage of epidemic outbreaks when the effects of
recovery and death can be ignored. The behaviors of the SI model
are not only of theoretical interest, but also of practical
significance beyond the physics community. However, this has not
been carefully investigated thus far.

Very recently, Barth\'elemy \emph{et al.}
\cite{Barthelemy2004,Barthelemy2005} studied the SI model in
Barab\'asi-Albert (BA) scale-free networks \cite{Barabasi1999},
and found that the density of infected nodes, denoted by $i(t)$,
grows approximately in the exponential form, $i(t)\sim e^{ct}$,
where the time scale $c$ is proportional to the ratio between the
second and the first moments of the degree distribution, $c\sim
\langle k^2 \rangle/ \langle k \rangle$ . Since the degree
distribution of the BA model obeys the power-law form $P(k)\sim
k^{-\gamma}$ with $\gamma=3$, this epidemic process has an
infinite spreading velocity in the limit of infinite population.
Following a similar process on \emph{random Apollonian networks}
\cite{Zhou2005,Gu2005,Zhang2005} and the
Barrat-Barth\'elemy-Vespignani networks
\cite{Barrat2004a,Barrat2004b}, Zhou \emph{et al.} investigated
the effects of clustering \cite{Zhou2005} and weight distribution
\cite{Yan2005} on SI epidemics. And by using the theory of
branching processes, Vazquez obtained a more accurate solution of
$i(t)$, including the behaviors with large $t$ \cite{Vazquez2006}.
The common assumption in all the aforementioned works
\cite{Barthelemy2004,Barthelemy2005,Zhou2005,Yan2005} is that each
node's potential infection-activity (infectivity), measured by its
possibly maximal contribution to the propagation process within
one time step, is strictly equal to its degree. Actually, only the
contacts between susceptible and infected nodes have possible
contributions in epidemic processes. However, since in a real
epidemic process, an infected node usually does not know whether
its neighbors are infected, the standard network SI model assumes
that each infected node will contact every neighbor once within
one time step \cite{Barthelemy2004}, thus the infectivity is equal
to the node degree.

The node with very large degree is called a \emph{hub} in network
science
\cite{Albert2002,Dorogovtsev2002,Newman2003,Boccaletti2006}, while
the node with great infectivity in an epidemic contact network is
named \emph{superspreader} in the epidemiological literature
\cite{Bassetti2005,Small2005,Bai2006}. All the previous studies on
SI network model have a basic assumption, that is, $hub \equiv
superspreader$. This assumption is valid in some cases where the
hub node is much more powerful than the others. However, there are
still many real spreading processes, which can not be properly
described by this assumption. Some typical examples are as
follows.

$\bullet$ In the broadcasting process, the forwarding capacity of
each node is limited. Especially, in wireless multihop ad hoc
networks, each node usually has the same power thus almost the
same forwarding capacity \cite{Gupta2000}.

$\bullet$ In epidemic contact networks, the hub node has many
acquaintances; however, he/she could not contact all his/her
acquaintances within one time step. Analogously, although a few
individuals have hundreds of sexual partners, their sexual
activities are not far beyond a normal level due to the
physiological limitations
\cite{Liljeros2001,Liljeros2003,Schneeberger2004}.

$\bullet$ In some email service systems, such as the Gmail system
schemed out by Google \cite{Gmail}, one can be a client only if
he/she received at least one invitation from some existing
clients. And after he/she becomes a client, he/she will have the
ability to invite others. However, the maximal number of
invitations he/she can send per a certain period of time is
limited.

$\bullet$ In network marketing processes, the referral of a
product to potential consumers costs money and time (e.g. a
salesman has to make phone calls to persuade his social
surrounding to buy the product). Thus, generally speaking, the
salesman will not make referrals to all his acquaintances
\cite{Kim2006}.

In addition, since the infectivity of each node is assigned to be
equal to its degree, one cannot be sure which (the power-law
degree distribution, the power-law infectivity distribution, or
both) is the main reason that leads to the virtually infinite
propagation velocity of the infection.

\begin{figure}
\scalebox{0.8}[0.8]{\includegraphics{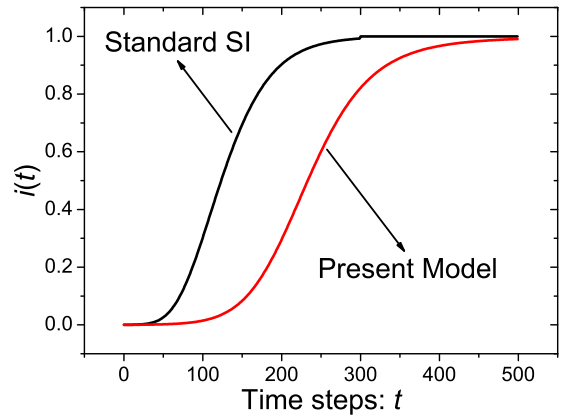}} \caption{(Color
online) The infected density $i(t)$ vs time, where $i(t)=I(t)/N$.
The black and red curves result from the standard SI network model
and the present model. The numerical simulations are implemented
based on the BA network \cite{Barabasi1999} of size $N=5000$ and
with average degree $\langle k\rangle =6$. The spreading rate is
given as $\lambda=0.01$, and the data are averaged over 5000
independent runs.}
\end{figure}

\begin{figure}
\scalebox{0.8}[0.8]{\includegraphics{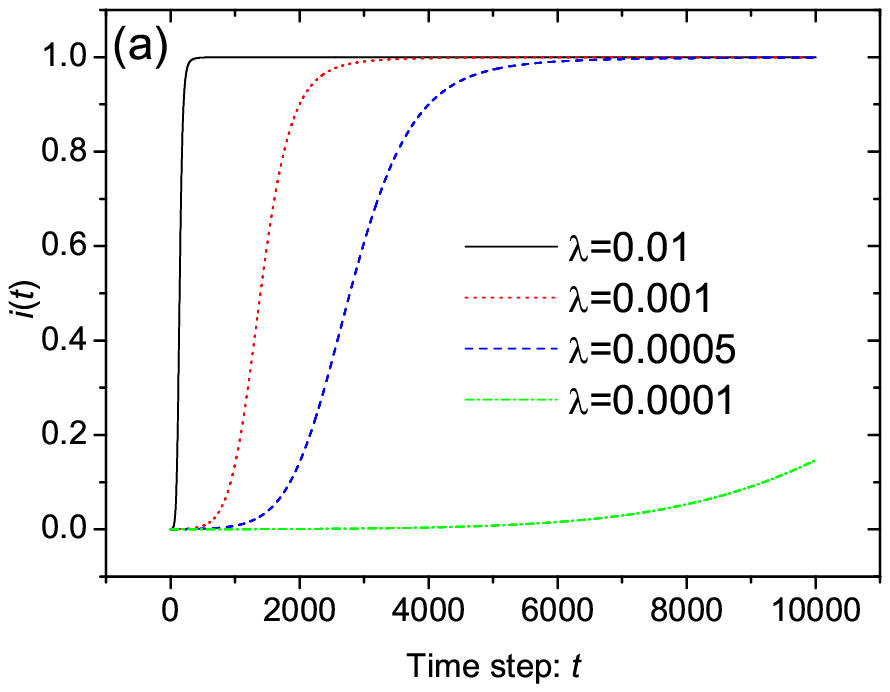}}
\scalebox{0.8}[0.8]{\includegraphics{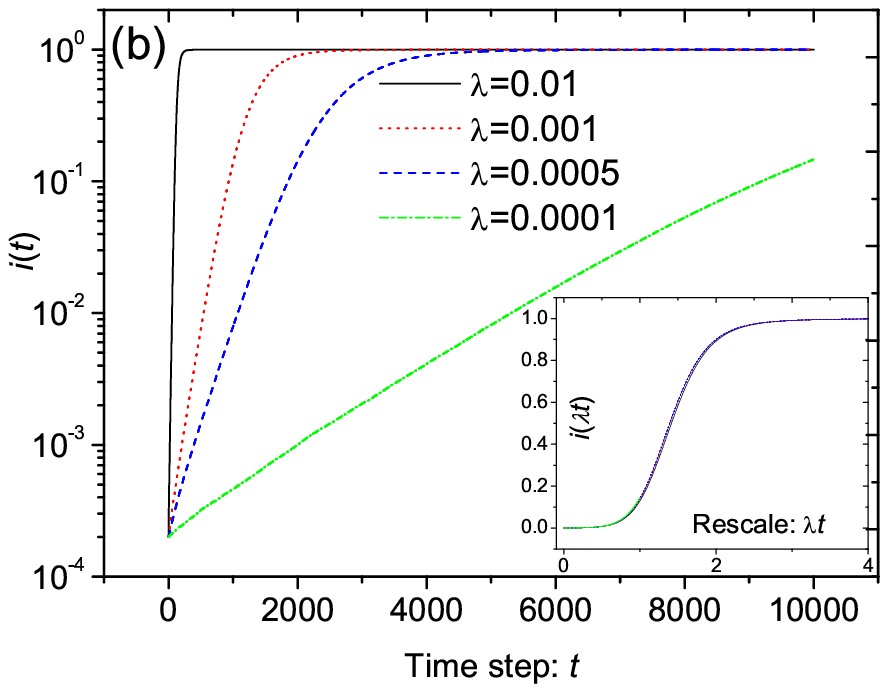}} \caption{(Color
online) The infected density $i(t)$ vs time in normal (a) and
single-log (b) plots. The black solid, red dot, green dash and
blue dash-dot curves correspond to $\lambda=0.01, 0.001, 0.0005$
and 0.0001, respectively. In single-log plot (b), the early
behavior of $i(t)$ can be well fitted by a straight line,
indicating the exponential growth of infected population. The
inset shows the rescaled curves $i(\lambda t)$. The four curves
for different $\lambda$ collapse to one curve in the new scale
$\lambda t$. The numerical simulations are implemented based on a
BA network of size $N=5000$ and with average degree $\langle
k\rangle =6$, and the data are averaged over 5000 independent
runs.}
\end{figure}

\section{Model}
Different from the previous works, here we investigate the SI
process on scale-free networks with identical infectivity. In our
model, individuals can be in two discrete states, either
susceptible or infected. The total population (i.e. the network
size) $N$ is assumed to be constant; thus, if $S(t)$ and $I(t)$
are the numbers of susceptible and infected individuals at time
$t$, respectively, then
\begin{equation}
N=S(t)+I(t).
\end{equation}
Denote by $\lambda$ the \emph{spreading rate} at which each
susceptible individual acquires infection from an infected
neighbor during one time step. Accordingly, one can easily obtain
the probability that a susceptible individual $x$ will be infected
at time step $t$ to be
\begin{equation}
\lambda_x(t)=1-(1-\lambda)^{\theta(x,t-1)},
\end{equation}
where $\theta(x,t-1)$ denotes the number of contacts between $x$
and the infected individuals at time $t-1$. For small $\lambda$,
one has
\begin{equation}
\lambda_x(t)\approx \lambda \theta(x,t-1).
\end{equation}

In the standard SI network model
\cite{Barthelemy2004,Barthelemy2005,Zhou2005}, each infected
individual will contact all its neighbors once at each time step,
thus the infectivity of each node is defined by its degree and
$\theta(x,t)$ is equal to the number of its infected neighbors at
time $t$. In the present model, we assume every individual has the
same infectivity $A$, in which, at every time step, each infected
individual will generate $A$ contacts where $A$ is a constant.
Multiple contacts to one neighbor are allowed, and contacts
between two infected ones, although having no effect on the
epidemic dynamics, are also counted just like the standard SI
model. The dynamical process starts by selecting one node
randomly, assuming it is infected.

\section{Spreading Velocity}
In the standard SI network model, the average infectivity equals
the average degree $\langle k \rangle$. Therefore, in order to
compare the proposed model with the standard one, we set
$A=\langle k \rangle$. As shown in Fig. 1, the dynamical behaviors
of the present model and the standard one are clearly different:
The velocity of the present model is much less than that of the
standard model.

In the following discussions, we focus on the proposed model.
Without loss of generality, we set $A=1$. Denote by $i_k(t)$ the
density of infected $k$-degree nodes. Based on the mean-field
approximation, one has
\begin{equation}
\frac{\texttt{d}i_k(t)}{\texttt{d}t}=\lambda
k[1-i_k(t)]\sum_{k'}\frac{1}{k'}\frac{k'P(k')i_{k'}(t)}{\sum_{k''}k''P(k'')},
\end{equation}
where $P(k)$ denotes the probability that a randomly selected node
has degree $k$. The factor $\frac{1}{k'}$ accounts for the
probability that one of the infected neighbors of a node, with
degree $k'$, will contact this node at the present time step. Note
that the infected density is given by
\begin{equation}
i(t)=\sum_ki_k(t)P(k),
\end{equation}
so Eq. (4) can be rewritten as
\begin{equation}
\frac{\texttt{d}i_k(t)}{\texttt{d}t}=\frac{\lambda k}{\langle k
\rangle}[1-i_k(t)]i(t).
\end{equation}
Manipulating the operator $\sum_kP(k)$ on both sides, and
neglecting terms of order $\mathbb{O}(i^2)$, one obtains the
evolution behavior of $i(t)$ as follows:
\begin{equation}
i(t)\sim \texttt{e}^{ct},
\end{equation}
where $c\propto \lambda$ is a constant independent of the
power-law exponent $\gamma$.

\begin{figure}
\scalebox{0.8}[0.8]{\includegraphics{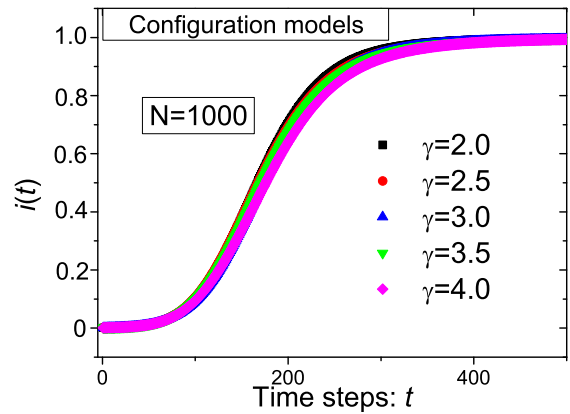}} \caption{(Color
online) The infected density $i(t)$ vs time for different
$\gamma$. The black squares, red circles, blue up-triangles, green
down-triangles, and pink diamonds (from up to down) denote the
cases of $\gamma=2.0,2.5,3.0,3.5$ and 4.0, respectively. The
numerical simulations are implemented based on the scale-free
configuration network model. The networks are of size $N=1000$ and
with average degree $\langle k\rangle =6$, the spreading rate is
given as $\lambda=0.01$, and the data are averaged over 10000
independent runs.}
\end{figure}

\begin{figure}
\scalebox{0.42}[0.42]{\includegraphics{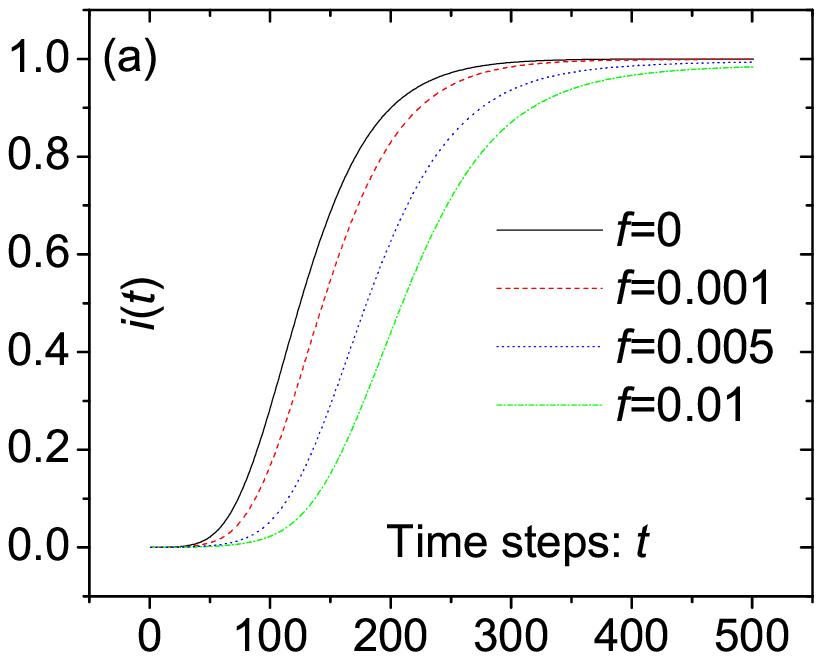}}
\scalebox{0.42}[0.42]{\includegraphics{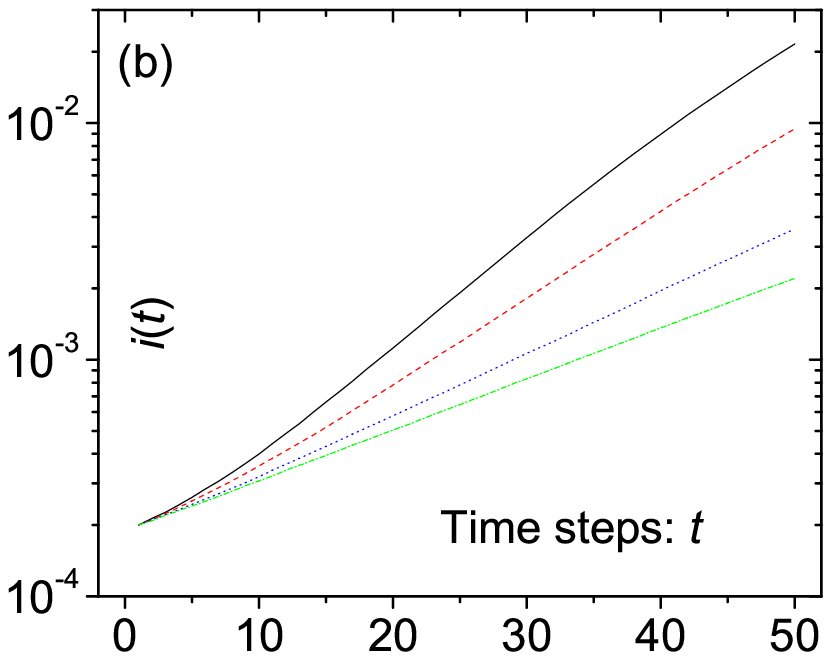}}
\scalebox{0.42}[0.42]{\includegraphics{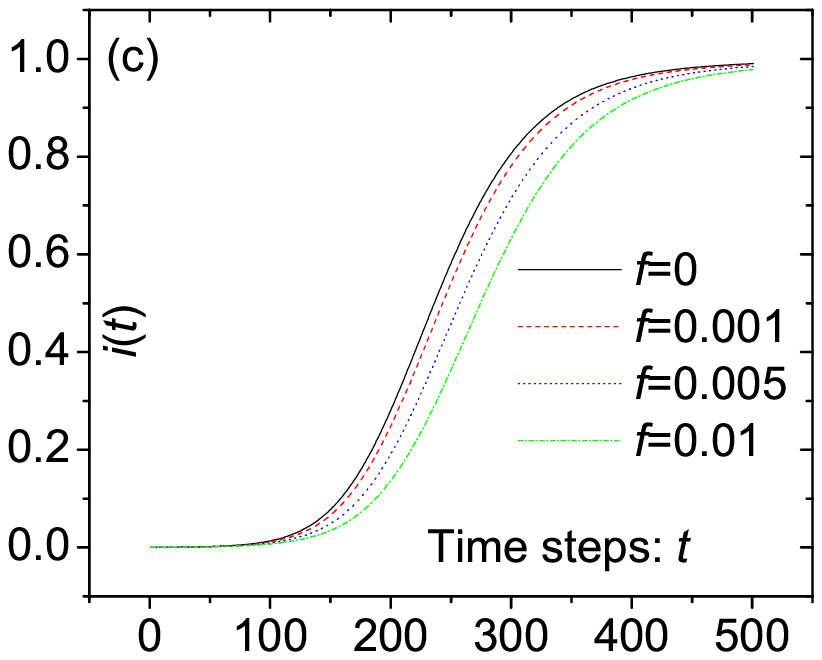}}
\scalebox{0.42}[0.42]{\includegraphics{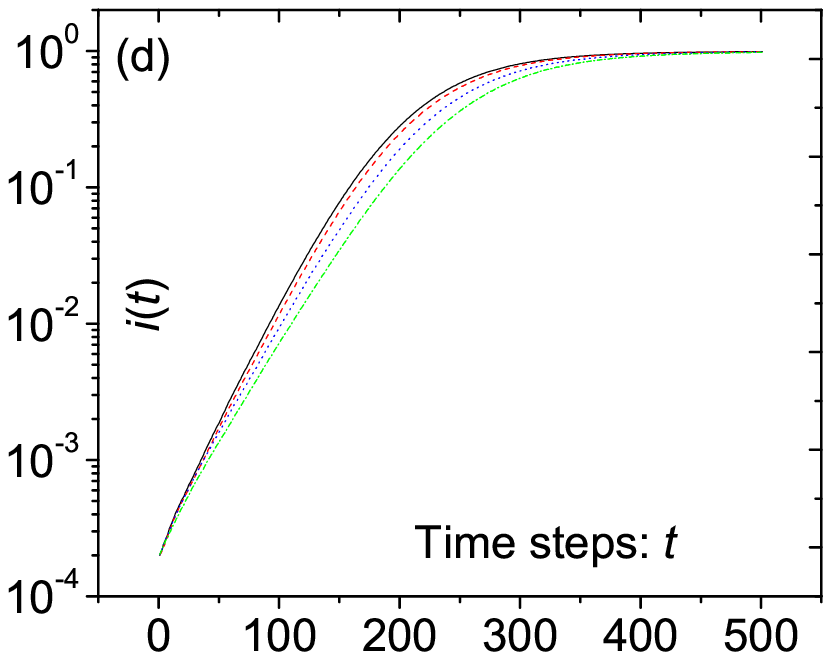}} \caption{(Color
online) The infected density $i(t)$ va time with different
vaccinating ranges. Figure 4a and 4b show the results of targeted
immunization for the standard SI process in normal and single-log
plots, respectively. Correspondingly, figure 4c and 4d display the
results for the present model. In all the four panels, the black
solid, red dash, blue dot and green dash-dot curves represent the
cases of $f=0$, 0.001, 0.005 and 0.01, respectively. The numerical
simulations are implemented based on a BA network of size $N=5000$
and with average degree $\langle k\rangle =6$, the spreading rate
is given as $\lambda=0.01$, and the data are averaged over 5000
independent runs. For comparison, the infectivity of the present
model is set as $A=\langle k\rangle =6$.}
\end{figure}

\begin{figure}
\scalebox{0.8}[0.8]{\includegraphics{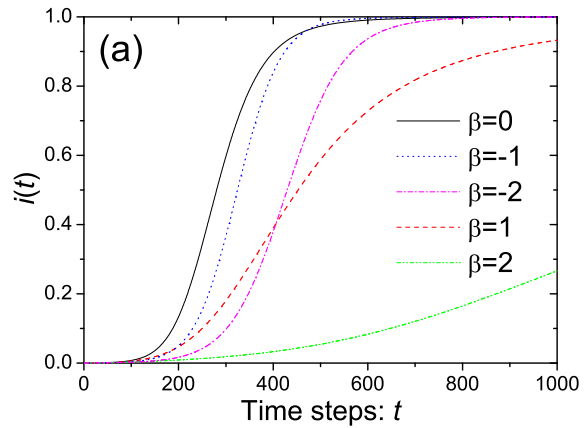}}
\scalebox{0.8}[0.8]{\includegraphics{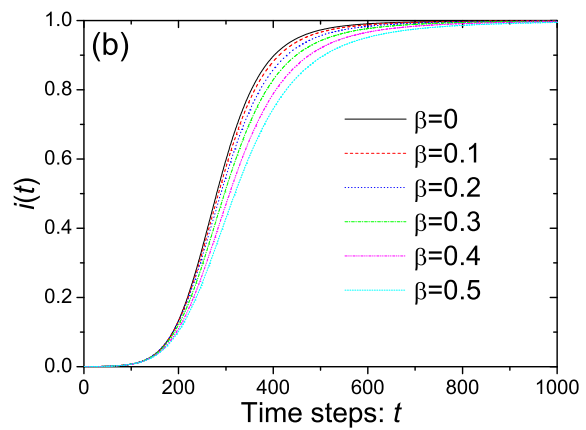}}
\scalebox{0.8}[0.8]{\includegraphics{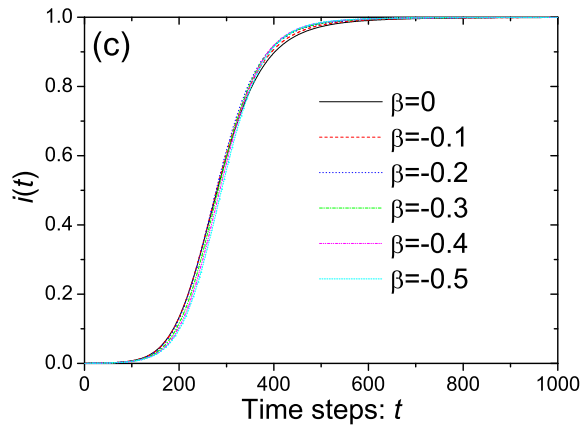}} \caption{(Color
online) The infected density $i(t)$ vs time for different $\beta$.
In Figure 5(a), the black solid, blue dot, magenta dash-dot, red
dash and green dash-dot-dot curves correspond to $\beta=0$, -1,
-2, 1 and 2, respectively. In Figure 5(b), the black solid, red
dash, blue dot, green dash-dot, magenta dash-dot-dot and cyan
short-dash curves, from up to down, correspond to $\beta=0$, 0.1,
0.2, 0.3, 0.4 and 0.5, respectively. In Figure 5(c), the black
solid, red dash, blue dot, green dash-dot, magenta dash-dot-dot
and cyan short-dash curves correspond to $\beta=0$, -0.1, -0.2,
-0.3, -0.4 and -0.5, respectively. The numerical simulations are
implemented based on the extensional BA network of size $N=5000$
and with average degree $\langle k\rangle =6$, the spreading rate
is given as $\lambda=0.01$ and the data are averaged over 5000
independent runs.}
\end{figure}

In Fig. 2, we report the simulation results of the present model
for different spreading rates ranging from 0.0001 to 0.01. The
curves $i(t)$ vs $t$ can be well fitted by a straight line in
single-log plot for small $t$ with slope proportional to $\lambda$
(see also the inset of Fig. 2b, where the curves for different
values of $\lambda$ collapse to one curve in the time scale
$\lambda t$), which strongly supports the analytical results.
Furthermore, based on the scale-free configuration model
\cite{Newman2001,Chung2002}, we investigated the effect of network
structure on epidemic behaviors. Different from the standard SI
network model \cite{Barthelemy2004,Barthelemy2005}, which is
highly affected by the power-law exponent $\gamma$, as shown in
Fig. 3, the exponent $\gamma$ here has almost no effects on the
epidemic behaviors of the present model. In other words, in the
present model, the spreading rate $\lambda$, rather than the
heterogeneity of degree distribution, governs the epidemic
behaviors.

\section{Targeted Immunization}
An interesting and practical problem is whether the epidemic
propagation can be effectively controlled by vaccination aiming at
part of the population \cite{Pastor2003,Zhou2006,Li2006}. The most
simple case is to select some nodes completely randomly, and then
vaccinate them. By applying the percolation theory, this case can
be exactly solved \cite{Cohen2000,Callway2000}. The corresponding
result shows that it is not an efficient immunization strategy for
highly heterogeneous networks such as scale-free networks.
Recently, some efficient immunization strategies for scale-free
networks are proposed. On the one hand, if the degree of each node
can not be known clearly, an efficient strategy is to vaccinate
the random neighbors of some randomly selected nodes since the
node with larger degree has greater chance to be chosen by this
double-random chain than the one with small degree
\cite{Huerta2002,Cohen2003}. On the other hand, if the degree of
each node is known, the most efficient immunization strategy is
the so-called \emph{targeted immunization}
\cite{Pastor2002,Madar2004}, wherein the nodes of highest degree
are selected to be vaccinated (see also a similar method in Ref.
\cite{Dezso2002}).

Here, we compare the performance of the targeted immunization for
standard SI model and the present model. To implement this
immunization strategy, a fraction of population having highest
degree, denoted by $f$, are selected to be vaccinated. That is to
say, these $Nf$ nodes will never be infected but the contacts
between them and the infected nodes are also counted. Clearly, in
both the two models, the hub nodes have more chances to receive
contacts from their infected neighbors, thus this targeted
immunization strategy must slow down the spreading velocity. In
Fig. 4a and Fig. 4b, we report the simulation results for the
standard SI model. The spreading velocity remarkably decreases
even only a small fraction, $f=0.001$, of population get
vaccinated, which strongly indicate the efficiency of the targeted
immunization. Relatively, the effect of the targeted immunization
for the present model is much weaker (see Fig. 4c and Fig. 4d).
The difference is more obvious in the single-log plot (see Fig. 4b
and Fig. 4d): The slope of the curve $i(t)\sim t$, which denotes
the time scale of the exponential term that governs the epidemic
behaviors, sharply decreases even only a small amount of hub nodes
are vaccinated in standard SI process while changes slightly in
the present model.

\section{Fast Spreading Strategy}
As mentioned in the Sec. 4, previous studies about network
epidemic processes focus on how to control the epidemic spreading,
especially for scale-free networks. Contrarily, few studies aim at
accelerating the epidemic spreading process. However, a fast
spreading strategy may be very useful for enhancing the efficiency
of network broadcasting or for making profits from network
marketing. In this section, we give a primary discussion on this
issue by introducing and investigating a simple fast spreading
strategy. Since the whole knowledge of network structure may be
unavailable for large-scale networks, here we assume only local
information is available.

In our strategy, at every time step, each infected node $x$ will
contact its neighbor $y$ (in the broadcasting process, it means to
forward a message to node $y$) at a probability proportional to
$k_y^\beta$, where $k_y$ denotes the degree of $y$. There are two
ingredients simultaneously affect the performance of the present
strategy. On the one hand, the strategy preferring large-degree
node (i.e. the strategy with $\beta>0$) corresponds to shorter
average distance in the path searching algorithm
\cite{Adamic2001,Kim2002}, thus it may lead to faster spreading.
On the other hand, to contact an already infected node (i.e. to
forward a message to a node having already received this message)
has no effects on the spreading process, and the nodes with larger
degrees are more easily to be infected according to Eq. (6) in the
case of $\beta=0$. Therefore, the strategy with $\beta>0$ will
bring many redundant contacts that may slow down the spreading.
For simplicity, we call the former the \emph{shorter path effect}
(SPE), and the latter the \emph{redundant contact effect} (RCE).

Figure 5(a) shows the density of infected individuals $i(t)$ as a
function of $t$ for different $\beta$. Clearly, due to the
competition between the two ingredients, SPE and RCE, the
strategies with too large (e.g. $\beta=1,2$) or too small (e.g.
$\beta=-1,-2$) $\beta$ are inefficient comparing with the unbiased
one with $\beta=0$. The cases when $\beta$ is around zero are
shown in Figs. 5(b) and 5(c). In Fig. 5(b), one can see that the
RCE plays the major role in determining the epidemic velocity when
$\beta>0$; that is, larger $\beta$ leads to slower spreading. As
shown in Fig. 5(c), the condition is much more complex when
$\beta<0$: In the early stage, the unbiased strategy seems better;
however, as time goes on, it is exceeded by the others.

\section{Conclusion and Discussion}
Almost all the previous studies about the SI model in scale-free
networks essentially assume that the nodes of large degrees are
not only dominant in topology, but also the superspreaders.
However, not all the SI network processes can be appropriately
described under this assumption. Typical examples include the
network broadcasting process with a limited forwarding capacity,
the epidemics of sexually transmitted diseases where all
individuals' sexual activities are pretty much the same due to the
physiological limitations, the email service systems with limited
ability to accept new clients, the network marketing systems where
the referral of products to potential consumers costs money and
time, and so on. Inspired by these practical requirements, in this
article we have studied the behaviors of susceptible-infected
epidemics on scale-free networks with identical infectivity. The
infected population grows in an exponential form in the early
stage. However, different from the standard SI network model, the
epidemic behavior is not sensitive to the power-law exponent
$\gamma$, but is governed only by the spreading rate $\lambda$.
Both the simulation and analytical results indicate that it is the
heterogeneity of infectivities, rather than the heterogeneity of
degrees, governs the epidemic behaviors. Further more, we compare
the performances of targeted immunization on the standard SI
process and the present model. In this standard SI process, the
spreading velocity decreases remarkably even only a slight
fraction of population are vaccinated. However, since the
infectivity of the hub nodes in the present model is just equal to
that od the small-degree node, the targeted immunization for the
present model is much less efficient.

We have also investigated a fast spreading strategy when only
local information is available. Different from previous reports
about some relative processes taking place on scale-free networks
\cite{Adamic2001,Kim2002}, we found that the strategy preferring
small-degree nodes is more efficient than those preferring large
nodes. This result indicates that the redundant contact effect is
more important than the shorter path effect. This finding may be
useful in practice. Very recently, some authors suggested using a
quantity named \emph{saturation time} to estimate the epidemic
efficiency \cite{Zhu2004,Saramaki2005}, which means the time when
the infected density, $i(t)$, firstly exceeds 0.9. Under this
criterion, the optimal value of $\beta$ leading to the shortest
saturation time is -0.3.

Some recent studies on network traffic dynamics show that the
networks will have larger throughput if using routing strategies
preferring small-degree nodes \cite{Yin2006,Wang2006,Yan2006}. It
is because this strategy can avoid possible congestion occurring
at large-degree nodes. Although the quantitative results are far
different, there may exist some common features between network
traffic and network epidemic. We believe that our work can further
enlighten the readers on this interesting subject.

\begin{acknowledgments}
This work was partially supported by the National Natural Science
Foundation of China under Grant Nos. 70471033, 10472116, 10532060,
70571074 and 10547004, the Special Research Founds for Theoretical Physics Frontier Problems
under Grant No. A0524701, and Specialized Program under the
Presidential Funds of the Chinese Academy of Science.
\end{acknowledgments}

\end{document}